\documentclass[12pt]{article}
\usepackage{sw20lart}

\input tcilatex
\QQQ{Language}{
British English
}

\begin{document}

\section{Introduction}

Both the matrix eigenvalue problem and the inverse eigenvalue problem are
the subjects of intensive research [1,2]. This paper addresses both
problems. A means of producing matrices with any specified eigenvalues and
whose eigenvectors are known is presented. Both the elements of the matrix
and the eigenvectors of the matrix are given in analytic form; this should
facilitate the use of such matrices for analysis in other areas of linear
algebra. A reversal of the matrix-generation procedure naturally suggests a
matrix-diagonalization algorithm, and we derive the equation which needs to
be solved to realize this. The findings we are going to present grew out of
research into quantum foundations[3], but are here presented from a purely
mathematical point of view.

The organization of this paper is as follows. In the next section, we
present the basic result that makes possible the production of matrices with
any desired eigenvalues. In Section 3, we outline the diagonalization
routine that arises from it. In Section 4, we give the functions which for
the case $N=5$ enable the practical realization of both these projects. The
results of preliminary application of this theory are reported in Section 5.
We conclude with brief comments in Section 6.

\section{Matrix Generation}

\ Suppose that the $N^2$ functions $\phi (B_i;C_j)$ possess the
orthonormality property

\begin{equation}
\sum_{l=1}^N\phi (B_l;C_i)\phi ^{*}(B_l;C_j)=\delta _{ij},  \label{on1}
\end{equation}
where $B$ and $C$ are parameters that take the values $B_1$,$B_2$,...,$B_N$
and $C_1$, $C_2$,...,$C_N$ respectively. The quantities $\phi (B_i;C_j)$ are
functions of some variables, collectively denoted by $x$. If with the aid of
the $N$ arbitrary numbers $\lambda _1,\lambda _2,...\lambda _N$ we form the $%
N\times N$ matrix $[M]$ using the rule

\begin{equation}
M_{ij}=\sum_{n=1}^N\phi ^{*}(B_i;C_n)\phi (B_j;C_n)\lambda _n  \label{ei8}
\end{equation}
we find that the eigenvectors of this matrix are

\begin{equation}
\lbrack \xi _i]=\left( 
\begin{array}{c}
\phi ^{*}(B_1;C_i) \\ 
\phi ^{*}(B_2;C_i) \\ 
: \\ 
\phi ^{*}(B_N;C_i)
\end{array}
\right)  \label{ni9}
\end{equation}
with the respective eigenvalues $\lambda _i.$ Indeed if we write\ 
\begin{equation}
\lbrack V]=[M][\xi _i],  \label{te10}
\end{equation}
we find that the $k$th row of $[M][\xi _i]$ is

\begin{eqnarray}
V_k &=&\sum_{l=1}^NM_{kl}\phi ^{*}(B_l;C_i)  \nonumber \\
\ &=&\sum_{l=1}^N\left( \sum_{n=1}^N\phi ^{*}(B_k;C_n)\phi (B_l;C_n)\lambda
_n\right) \phi ^{*}(B_l;C_i)  \nonumber \\
\ &=&\sum_{n=1}^N\lambda _n\phi ^{*}(B_k;C_n)\sum_{l=1}^N\phi (B_l;C_n)\phi
^{*}(B_l;C_i)  \nonumber \\
\ &=&\sum_{n=1}^N\lambda _n\phi ^{*}(B_k;C_n)\delta _{ni}  \nonumber \\
\ &=&\lambda _i\phi ^{*}(B_k;C_i).  \label{el11}
\end{eqnarray}

Thus the eigenvalue equation 
\begin{equation}
\lbrack M][\xi _i]=\lambda _i[\xi _i]  \label{tw12}
\end{equation}
is satisfied. Property Eqn. (\ref{on1})\ means that the eigenvectors are
orthonormal: 
\begin{eqnarray}
\lbrack \xi _i]^{\dagger }[\xi _j] &=&\sum_l\phi (B_l;C_i)\phi ^{*}(B_l;C_j)
\nonumber \\
&=&\delta _{ij}.  \label{th13}
\end{eqnarray}

These results give us a means of generating matrices with specified
eigenvalues and known eigenvectors. We only need appropriate functions $\phi
(B_i;C_j).$ An example of such functions will be given in Section 4. Given
such functions, and a set of eigenvalues $\{\lambda _i\},$ we are able
generate an infinite number of different matrices by varying the argument $x$
of these functions. Each such matrix is generated together with its own
orthonormal set of eigenvectors.

On the other hand, given a value of $x$, we can generate an infinite number
of matrices by using different sets of eigenvalues. In this case, all these
matrices share the same set of orthonormal eigenvectors.

\section{A Diagonalization Algorithm}

Since each matrix is generated by means of a specific value of the argument $%
x,$ we should be able, at least in principle, to find a way of deducing this
value of $x$ for a given matrix. A procedure for doing this would constitute
a diagonalization routine. In this section, we derive one such method.

Suppose we are given a matrix $[M]$ formed by the prescription Eqn. (\ref
{ei8}). Its eigenvectors are given by Eqn. (\ref{ni9}). If we only knew the
value of $x$ that was used to form the matrix, then we could fix the
eigenvectors and recover the eigenvalues. In order to develop a way of
deducing the value of $x,$ let us start, for the sake of convenience, by
setting 
\begin{equation}
\xi (C_i;B_j)=\phi ^{*}(B_j;C_i)  \label{si16}
\end{equation}
so that

\begin{equation}
\lbrack \xi _i]=\left( 
\begin{array}{c}
\phi ^{*}(B_1;C_i) \\ 
\phi ^{*}(B_2;C_i) \\ 
: \\ 
\phi ^{*}(B_N;C_i)
\end{array}
\right) =\left( 
\begin{array}{c}
\xi (C_i,B_1) \\ 
\xi (C_i,B_2) \\ 
: \\ 
\xi (C_i,B_N)
\end{array}
\right) .  \label{fi15}
\end{equation}

Now, from

\begin{equation}
\lbrack M][\xi _i]=\lambda _i[\xi _i]  \label{si16}
\end{equation}
and the value of $x$ that was used to produce the matrix, we can recover the
eigenvalue $\lambda _i$ from the $r$th row by using

\begin{equation}
(\lambda _i)_r=\frac{\sum\nolimits_lM_{rl}\xi (C_i,B_l)}{\xi (C_i,B_r)}.
\label{se17}
\end{equation}

The $s$th row would give the same value. Hence, comparing the values from
the $r$th and $s$th rows, we can write 
\begin{equation}
(\lambda _i)_r-(\lambda _i)_s=0.\;\;  \label{ei18}
\end{equation}

This must be true for all possible combinations of $r$ and $s:$

\begin{equation}
\sum_{r=1}^{N-1}\sum_{s=r+1}^N\left[ (\lambda _i)_r-(\lambda _i)_s\right] =0.
\label{ni19}
\end{equation}

In other words:

\begin{equation}
\sum_{r=1}^{N-1}\sum_{s=r+1}^N\left[ \frac{\sum_lM_{rl}\xi (C_i,B_l)}{\xi
(C_i,B_r)}-\frac{\sum_lM_{sl}\xi (C_i,B_l)}{\xi (C_i,B_s)}\right] =0.
\label{tw20}
\end{equation}

But if the value of $x$ used is not correct, the equation is not satisfied.
Thus, this equation is a means of deducing the correct value of $x$. A form
of this equation better adapted to numerical work is

\begin{equation}
\sum_{i=1}^N\sum_{r=1}^{N-1}\sum_{s=r+1}^N\left[ \xi
(C_i,B_s)\sum_lM_{rl}\xi (C_i,B_l)-\xi (C_i,B_r)\sum_lM_{sl}\xi
(C_i,B_l)\right] =0.  \label{tw21}
\end{equation}

We have summed over all the eigenvalues by means of the index $i$ because we
must ensure that the equation is satisfied for every eigenvalue.

Once the correct value of $x$ $($denoted by $x_0)$ is found, the eigenvalues
are calculated using 
\begin{equation}
\left. (\lambda _i)_r=\frac{\sum_lM_{rl}\xi (C_i,B_l)}{\xi (C_i,B_r)}\right|
_{x_0}  \label{tw23}
\end{equation}
and the eigenvectors are 
\begin{equation}
\lbrack \xi _i]=\left. \left( 
\begin{array}{c}
\xi (C_i,B_1) \\ 
\xi (C_i,B_2) \\ 
: \\ 
\xi (C_i,B_N)
\end{array}
\right) \right| _{x_0}.  \label{tw24}
\end{equation}

As is wise in the solution of such equations, it is necessary to establish
that the value of $x$ obtained is not spurious by putting it back into the
equation. In this case, this means using it to calculate the eigenvalue
using every line of the eigenvector. Thus, for each eigenvalues, there
should be $N$ values, which should agree in order for this value of $x$ to
be acceptable. Of course, in the case where one of the elements of the
eigenvectors vanishes, it cannot be used to compute the eigenvalue, and
there will be fewer than $N$ values to compare.

\section{Probability Amplitudes\textbf{\ }}

In order to utilise the prescription for generating matrices with specified
eigenvalues and to realise the algorithm for diagonalizing them, we need the
functions $\phi (B_i;C_j).$ Since the origins of the method presented in
this paper lie in quantum theory, we shall call these functions probability
amplitudes. A source for such functions is spin theory in quantum
mechanics[3]. The treatment of a spin $s$ system yields functions suitable
for generating matrices of order $N=2s+1$. As an example, we take the spin-2
system. In the direction of the unit vector 
\begin{equation}
\widehat{\mathbf{c}}=(\sin \theta \cos \varphi ,\sin \theta \sin \varphi
,\cos \theta )  \label{tw25}
\end{equation}
its operator is[4]

\begin{equation}
\lbrack \widehat{\mathbf{c}}\cdot \mathbf{S}]=\left( 
\begin{array}{ccccc}
2\cos \theta & \sin \theta e^{-i\varphi } & 0 & 0 & 0 \\ 
\sin \theta e^{i\varphi } & \cos \theta & \frac{\sqrt{6}}2\sin \theta
e^{-i\varphi } & 0 & 0 \\ 
0 & \frac{\sqrt{6}}2\sin \theta e^{i\varphi } & 0 & -i\frac{\sqrt{6}}2\sin
\theta e^{-i\varphi } & 0 \\ 
0 & 0 & \frac{\sqrt{6}}2\sin \theta e^{i\varphi } & -\cos \theta & \sin
\theta e^{-i\varphi } \\ 
0 & 0 & 0 & \sin \theta e^{i\varphi } & -2\cos \theta
\end{array}
\right) .  \label{th39}
\end{equation}

The eigenvalues of this matrix are $2,1,0,-1$ and $-2$, with the respective
normalized eigenvectors

\begin{equation}
\lbrack \chi _{m=2}^{(\widehat{\mathbf{c}})}]=\left( 
\begin{array}{c}
\cos ^4\frac \theta 2e^{-i2\varphi } \\ 
\sin \theta \cos ^2\frac \theta 2e^{-i\varphi } \\ 
\frac{\sqrt{6}}4\sin ^2\theta \\ 
\sin \theta \sin ^2\frac \theta 2e^{i\varphi } \\ 
\sin ^4\frac \theta 2e^{i2\varphi }
\end{array}
\right) ,  \label{fo40}
\end{equation}

\begin{equation}
\lbrack \chi _{m=1}^{(\widehat{\mathbf{c}})}]=\left( 
\begin{array}{c}
\sin \theta \cos ^2\frac \theta 2e^{-i2\varphi } \\ 
(3\sin ^2\frac \theta 2-\cos ^2\frac \theta 2)\cos ^2\frac \theta
2e^{-i\varphi } \\ 
-\frac{\sqrt{6}}2\sin \theta \cos \theta \\ 
-(3\cos ^2\frac \theta 2-\sin ^2\frac \theta 2)\sin ^2\frac \theta
2e^{i\varphi } \\ 
-\sin \theta \sin ^2\frac \theta 2e^{i2\varphi }
\end{array}
\right) ,  \label{fo41}
\end{equation}

\begin{equation}
\lbrack \chi _{m=0}^{(\widehat{\mathbf{c}})}]=\left( 
\begin{array}{c}
\frac{\sqrt{6}}4\sin ^2\theta e^{-i2\varphi } \\ 
-\frac{\sqrt{6}}2\sin \theta \cos \theta e^{-i\varphi } \\ 
\frac 12(2\cos ^2\theta -\sin ^2\theta ) \\ 
\frac{\sqrt{6}}2\sin \theta \cos \theta e^{i\varphi } \\ 
\frac{\sqrt{6}}4\sin ^2\theta e^{i2\varphi }
\end{array}
\right) ,  \label{fo42}
\end{equation}
\begin{equation}
\lbrack \chi _{m=-1}^{(\widehat{\mathbf{c}})}]=\left( 
\begin{array}{c}
\sin \theta \sin ^2\frac \theta 2e^{-i2\varphi } \\ 
-(3\cos ^2\frac \theta 2-\sin ^2\frac \theta 2)\sin ^2\frac \theta
2e^{-i\varphi } \\ 
\frac{\sqrt{6}}2\sin \theta \cos \theta \\ 
(3\sin ^2\frac \theta 2-\cos ^2\frac \theta 2)\cos ^2\frac \theta
2e^{i\varphi } \\ 
-\sin \theta \cos ^2\frac \theta 2e^{i2\varphi }
\end{array}
\right)  \label{fo43}
\end{equation}
and

\begin{equation}
\lbrack \chi _{m=-2}^{(\widehat{\mathbf{c}})}]=\left( 
\begin{array}{c}
\sin ^4\frac \theta 2e^{-i2\varphi } \\ 
-\sin \theta \sin ^2\frac \theta 2e^{-i\varphi } \\ 
\frac{\sqrt{6}}4\sin ^2\theta \\ 
-\sin \theta \cos ^2\frac \theta 2e^{i\varphi } \\ 
\cos ^4\frac \theta 2e^{i2\varphi }
\end{array}
\right) .  \label{fo44}
\end{equation}

We have used the index $m$ to label the eigenvalues of the spin matrix
because that is the convention. Also, we need to keep the distinction
between the eigenvalues of this matrix and the eigenvalues $\{\lambda _i\}$
belonging to the matrix we wish to generate once we have obtained the
probability amplitudes.

For the direction 
\begin{equation}
\widehat{\mathbf{b}}=(\sin \theta ^{\prime }\cos \varphi ^{\prime },\sin
\theta ^{\prime }\sin \varphi ^{\prime },\cos \theta ^{\prime })
\label{fo44a}
\end{equation}
the eigenvectors of the operator are identical, except that $(\theta
,\varphi )$ is replaced by $(\theta ^{\prime },\varphi ^{\prime }).$ We now
associate the parameter $B$ with the direction $\widehat{\mathbf{b}}$, and
the parameter $C$ with the direction $\widehat{\mathbf{c}}$. The notation is
such that the values $B_1,B_2,...,B_5$ correspond respectively to the
eigenvalues $2,1,...,-2$ $.$ The same holds for $C$.

Some familiarity with spin theory in quantum mechanics is perhaps necessary
to fully appreciate these functions. Each element in each eigenvector is a
probability amplitude, and its modulus square is a probability for a certain
measurement[3]. However, from a mathematical point of view, all that is
needed is the fact that these elements possess the property Eqn. (\ref{on1}%
). The scalar products of the eigenvectors corresponding to $\widehat{%
\mathbf{b}}$ with those corresponding to $\widehat{\mathbf{c}}$ give the
most generalized forms of these functions. Thus the required quantities
are[5] 
\begin{equation}
\phi (B_i;C_j)=[\chi _{m_j}^{(\widehat{\mathbf{c}})}]^{\dagger }[\chi
_{m_i}^{(\widehat{\mathbf{b}})}].  \label{fo45}
\end{equation}
Evidently, in this case, $x=(\theta ,\varphi ,\theta ^{\prime },\varphi
^{\prime }).$

To illustrate the notation, we give a few of these functions:

\begin{eqnarray}
\phi (B_1;C_1) &=&[\chi _{m=2}^{(\widehat{\mathbf{c}})}]^{\dagger }[\chi
_{m=2}^{(\widehat{\mathbf{b}})}]  \nonumber \\
\ &=&\cos ^4\frac \theta 2\cos ^4\frac{\theta ^{\prime }}2e^{i2(\varphi
-\varphi ^{\prime })}  \nonumber \\
&&\ +\sin \theta ^{\prime }\sin \theta \cos ^2\frac{\theta ^{\prime }}2\cos
^2\frac \theta 2e^{i(\varphi -\varphi ^{\prime })}+\frac 38\sin ^2\theta
^{\prime }\sin ^2\theta  \nonumber \\
&&\ \ +\sin \theta ^{\prime }\sin \theta \sin ^2\frac{\theta ^{\prime }}%
2\sin ^2\frac \theta 2e^{-i(\varphi -\varphi ^{\prime })}  \nonumber \\
&&\ +\sin ^4\frac \theta 2\sin ^4\frac{\theta ^{\prime }}2e^{-i2(\varphi
-\varphi ^{\prime })},  \label{fo45a}
\end{eqnarray}
\begin{eqnarray}
\phi (B_1;C_2) &=&[\chi _{m=1}^{(\widehat{\mathbf{c}})}]^{\dagger }[\chi
_{m=2}^{(\widehat{\mathbf{b}})}]  \nonumber \\
\ &=&\sin \theta \cos ^4\frac{\theta ^{\prime }}2\cos ^2\frac \theta
2e^{i2(\varphi -\varphi ^{\prime })}  \nonumber \\
&&\ +(3\sin ^2\frac \theta 2-\cos ^2\frac \theta 2)\sin \theta ^{\prime
}\cos ^2\frac{\theta ^{\prime }}2\cos ^2\frac \theta 2e^{i(\varphi -\varphi
^{\prime })}  \nonumber \\
&&\ \ \ -\frac 34\sin ^2\theta ^{\prime }\sin \theta \cos \theta  \nonumber
\\
&&\ \ \ -(3\cos ^2\frac \theta 2-\sin ^2\frac \theta 2)\sin \theta ^{\prime
}\sin ^2\frac{\theta ^{\prime }}2\sin ^2\frac \theta 2e^{-i(\varphi -\varphi
^{\prime })}  \nonumber \\
&&\ -\sin \theta \sin ^4\frac{\theta ^{\prime }}2\sin ^2\frac \theta
2e^{-i2(\varphi -\varphi ^{\prime })}  \label{fo45b}
\end{eqnarray}
and 
\begin{eqnarray}
\phi (B_5;C_5) &=&[\chi _{m=-2}^{(\widehat{\mathbf{c}})}]^{\dagger }[\chi
_{m=-2}^{(\widehat{\mathbf{b}})}]  \nonumber \\
\ &=&\cos ^4\frac \theta 2\cos ^4\frac{\theta ^{\prime }}2e^{i2(\varphi
-\varphi ^{\prime })}  \nonumber \\
&&\ +\sin \theta ^{\prime }\sin \theta \sin ^2\frac{\theta ^{\prime }}2\sin
^2\frac \theta 2e^{i(\varphi -\varphi ^{\prime })}+\frac 38\sin ^2\theta
^{\prime }\sin ^2\theta  \nonumber \\
&&\ \ +\sin \theta ^{\prime }\sin \theta \cos ^2\frac{\theta ^{\prime }}%
2\cos ^2\frac \theta 2e^{-i(\varphi -\varphi ^{\prime })}  \nonumber \\
&&\ \ +\cos ^4\frac \theta 2\cos ^4\frac{\theta ^{\prime }}2e^{-i2(\varphi
-\varphi ^{\prime })}.  \label{fo45c}
\end{eqnarray}

There are 22 other functions. We remind ourselves that since 
\begin{equation}
\xi (C_i,B_j)=\phi ^{*}(B_j;C_i),  \label{fo45z}
\end{equation}
then 
\begin{equation}
\xi (C_i,B_j)=[\chi _{m_i}^{(\widehat{\mathbf{b}})}]^{\dagger }[\chi
_{m_j}^{(\widehat{\mathbf{c}})}].  \label{fo45y}
\end{equation}
These are the functions which enter into Eqn. (\ref{tw21}).

In general, the probability amplitudes satisfy the inter-dependence law

\begin{equation}
\phi (B_i;C_j)=\sum_{l=1}^5\phi (B_i;D_l)\phi ^{*}(D_l;C_j),  \label{fo45e}
\end{equation}
where $\widehat{\mathbf{d}}$ is a third direction to which corresponds the
new parameter $D$. This is a particular form of a fundamental quantum
property of three sets of probability amplitudes belonging to one quantum
system [6],

\begin{equation}
\psi (A_i;C_j)=\sum_{l=1}^N\chi (A_i;B_l)\phi (B_l;C_j).  \label{fo45f}
\end{equation}

In the context of quantum theory, the quantity $\left| \phi (B_i;C_j)\right|
^2$ is the probability that if the spin projection in the direction $%
\widehat{\mathbf{b}}$ is $m_i\hbar $, a measurement of it in the direction $%
\widehat{\mathbf{c}}$ yields the result $m_j\hbar $[3]. Here $\hbar $is
Planck's constant.

The parameters $B$ and $C$, as observed, are defined by the angles $(\theta
^{\prime },\varphi ^{\prime })$ and $(\theta ,\varphi ),$ and the indices of
each parameter correspond to the eigenvalues. When $(\theta ^{\prime
},\varphi ^{\prime })=(\theta ,\varphi )$ in Eqn. (\ref{fo45f}), $A=C$ and
Eqn. (\ref{on1}) is satisfied. Eqn. (\ref{on1}) is a special form of Eqn. (%
\ref{fo45f}). Within the context of quantum measurement theory, all this has
a ready interpretation[6]. However, we stress that it is not necessary to
understand this in order to use these functions to generate matrices
customized as to eigenvalues and type.

With these probability amplitudes, the eigenvector for the eigenvalue $%
\lambda _i$ is

\begin{equation}
\lbrack \xi _i]=\left( 
\begin{array}{c}
\phi ^{*}(B_1;C_i) \\ 
\phi ^{*}(B_2;C_i) \\ 
\phi ^{*}(B_3;C_i) \\ 
\phi ^{*}(B_4;C_i) \\ 
\phi ^{*}(B_5;C_i)
\end{array}
\right) =\left( 
\begin{array}{c}
\lbrack \chi _{m=2}^{(\widehat{\mathbf{b}})}]^{\dagger }[\chi _{m_i}^{(%
\widehat{\mathbf{c}})}] \\ 
\lbrack \chi _{m=1}^{(\widehat{\mathbf{b}})}]^{\dagger }[\chi _{m_i}^{(%
\widehat{\mathbf{c}})}] \\ 
\lbrack \chi _{m=0}^{(\widehat{\mathbf{b}})}]^{\dagger }[\chi _{m_i}^{(%
\widehat{\mathbf{c}})}] \\ 
\lbrack \chi _{m=-1}^{(\widehat{\mathbf{b}})}]^{\dagger }[\chi _{m_i}^{(%
\widehat{\mathbf{c}})}] \\ 
\lbrack \chi _{m=-2}^{(\widehat{\mathbf{b}})}]^{\dagger }[\chi _{m_i}^{(%
\widehat{\mathbf{c}})}]
\end{array}
\right) .  \label{fo46}
\end{equation}

Of course, since the matrix Eqn. (\ref{th39}) is a particular case of the
form of Eqn. (\ref{ei8}), its eigenvectors are given by Eqn. (\ref{fo46});
evidently, they correspond to the angles $\theta ^{\prime }=0$ and $\varphi
^{\prime }=0.$ In other words, for this case, the unit vector $\widehat{%
\mathbf{b}}$ is in the $z$ direction.

The association between the eigenvalues of the matrix we are generating and
their eigenvectors is through Eqn. (\ref{ei8}), where the value of $C$ and
the eigenvalue have the same index.

\section{Results}

\ A Fortran 77 program written to generate the matrices confirms that
different families of matrices are obtained depending on the values of $%
\theta ,\varphi ,\theta ^{\prime }$ and $\varphi ^{\prime },$and on the
character of the eigenvalues $\{\lambda _i\}.$ The following observations
are made, and can be predicted from the structure of the probability
amplitudes and of Eqn. (\ref{ei8}).

If all the angles are zero or if $\theta =\theta ^{\prime }$ and $\varphi
=\varphi ^{\prime },$ the resulting matrix is diagonal.

If the eigenvalues are real, the matrix is Hermitian, but its elements
depend on what values of the angles are used.

Whatever the eigenvalues, the matrix is symmetric if $\varphi =\varphi
^{\prime }.$

If $\varphi =\varphi ^{\prime },$ the eigenvectors are real.

If all the eigenvalues are pure imaginary, and the arguments are arbitrary,
the matrix is anti-Hermitian.

If all the eigenvalues are pure imaginary, and $\varphi =\varphi ^{\prime },$
the matrix is imaginary but symmetric.

If the eigenvalues are arbitrary and the values of the arguments also
arbitrary, the matrix is general.

The kinds of families that can be generated with different combinations of
the eigenvectors and the arguments have not been fully investigated, but it
seems probable that it does not require much ingenuity to generate such
special forms as tridiagonal matrices, etc. Clearly, with this method, much
of the inverse-eigenvalue problem is solved. Further investigation should
show how to chose the values of the angles so as to obtain the desired
matrices with specific values of certain elements.

As far as the diagonalization algorithm is concerned, it has only been
partially validated at this time. The main complication here is that the
non-linear equation (\ref{tw21}) is in four variables. Therefore, solving it
is somewhat complicated. The following was however achieved. A matrix was
generated by means of Eqn. (\ref{ei8}). In order to diagonalize it, three of
the arguments were held at their actual values, and the remaining one was
determined with the use of Eqn. (\ref{tw21}). Which one of the variables to
treat as unknown could be decided at pleasure. The value of this variable
was determined from Eqn. (\ref{tw21}) by means of the bisection method. That
this approach was successful indicates that once a fast and reliable method
of solving a non-linear equation in four variables is employed, the
algorithm will prove to be of some utility.

\section{Conclusion and Discussion}

In this paper, we have presented a prescription for forming matrices in such
a way that their eigenvalues and eigenvectors are known. The method is very
general indeed, and simply by varying the arguments, different kinds of
matrices can be obtained. Always, the normalised eigenvectors are
simultaneously given. The eigenvectors can be chosen to be real if desired,
and the any combination of eigenvalues can be used. By essentially reversing
the matrix-generation procedure, we have proposed an algorithm for
diagonalizing matrices formed this way.

A good amount of work is still needed in order to fully understand and
utilize the methods presented. For example, it is necessary to classify more
properly and completely according to the values of the arguments $\theta
,\varphi ,\theta ^{\prime }$ and $\varphi ^{\prime }$ the kinds of matrices
that can be generated. Such information would be particularly helpful in the
solution of the non-linear equation. If a matrix is such that one or more of
the arguments must have certain values, this makes so much easier the job of
solving the equation, since the number of variables is reduced.

The matrices dealt with here are of order $5$. In order to deal with
matrices of higher order, it is necessary to have the functions $\{\phi \}$
for those cases. A source of these functions will always be spin theory, but
treatment of other $N$-dimensional quantum systems should produce other
forms of the functions. Each such set of functions probably produces
matrices of different characters. As such, it expands the range of uses to
which we can put these matrices.

\textbf{Acknowledgements}

The author thanks the Staff and the Director at the Institute for
Mathematical Sciences in Chennai, India, where this work was carried out.

\section{References}

1. G. H. Golub, H. A. van der Vorst, Eigenvalue computation in the 20th
century, J. Comp. and Appl. Math. 123 (2000) 35-36

2. B. Boley, G. H. Golub, A survey of matrix inverse eigenvalue problems,
Inverse Problems 3 (1987) 595-622

3. H. V. Mweene, Derivation of spin vectors and operators from first
principles, quant-ph/9905012

4. See any standard text on quantum mechanics, such as B. H. Bransden and C.
J. Joachain, Introduction to Quantum Mechanics, Longman Scientific \&
Technical, 1989.

5. H. V. Mweene, Generalized probability amplitudes for spin projection
measurements on spin 2 systems, quant-ph/0502005

6. A. Land\'e, New Foundations of Quantum Mechanics, Cambridge University
Press, 1965.

\end{document}